\documentclass[aps,pre,twocolumn,showpacs,superscriptaddress,nofootinbib]{revtex4-2}

\usepackage{amsmath,amssymb,amsfonts}
\usepackage{bm}
\usepackage{graphicx}
\usepackage{hyperref}
\usepackage{physics}
\usepackage[mathscr]{eucal}

\begin{document}

\title{Nonlinear Coherent Transport in 2D Thermal Metamaterials: From Solitons and Topological Defects to Quantum Computing}

\author{R. A. C. Correa}
\email{rafael.a.correa@capgemini.com}
\affiliation{Capgemini, São Paulo Corporate Towers, Av. Pres. Juscelino Kubitschek 1909, São Paulo, SP 04543-907, Brazil}

\author{K. N. M. Sharma}
\email{sharmamadhavwork@gmail.com}
\affiliation{Capgemini Quantum Lab, Place de l'Étoile, 11 rue de Tilsitt, 75017 Paris, France}
\affiliation{Univ. Grenoble Alpes, CNRS, LPMMC, 38000 Grenoble, France}

\author{P. Lolur}
\email{phalgun.lolur@capgemini.com}
\affiliation{Capgemini Quantum Lab, Place de l'Étoile, 11 rue de Tilsitt, 75017 Paris, France}

\author{J. van Velzen}
\email{julian.van.velzen@capgemini.com}
\affiliation{Capgemini Quantum Lab, Place de l'Étoile, 11 rue de Tilsitt, 75017 Paris, France}
\date{\today}

\begin{abstract}
Understanding heat transport in low-dimensional and nano-architectured materials remains a central challenge in nonequilibrium statistical physics due to persistent deviations from Fourier's law. These deviations are driven by anharmonicity, reduced dimensionality, and the emergence of long-lived coherent excitations. In this work, we develop a unified theoretical framework for two-dimensional thermal metamaterials that combines nonlinear lattice dynamics, soliton-based effective field theories, and geometrically organized defect networks as guiding structures for energy flow. We introduce minimal discrete and continuum-inspired models suitable for controlled benchmarking of thermal transport in patterned two-dimensional architectures and identify a two-channel transport mechanism in which coherent nonlinear excitations coexist with incoherent hydrodynamic modes. The interplay between these channels is shown to be highly sensitive to geometry, nonlinearity, and temperature, offering new avenues for thermal management. We establish rigorous connections between microscopic nonlinearity, geometry-driven channeling of heat in two dimensions, and quantum-enabled exploration of both high-occupation classical regimes and genuinely quantum regimes beyond the reach of standard simulation strategies. The theoretical predictions are corroborated by recent experimental and computational results in Stone-Wales-defected PdSSe monolayers and silicon phononic crystal nanostructures, which exhibit ultra-low thermal conductivity coexisting with high carrier mobility and strong anisotropy---direct manifestations of the two-channel mechanism. This synthesis provides actionable guidance for the design of engineered heat-spreading architectures and positions quantum simulation as a transformative tool for advancing the theory of nonlinear heat transport.
\end{abstract}

\maketitle

\section{Introduction}

Heat transport in low-dimensional systems has remained a central problem in nonequilibrium statistical physics for several decades \cite{Lepri2003, Dhar2008}. In contrast to bulk three-dimensional solids, where Fourier's law of heat conduction provides an accurate macroscopic description, one- and two-dimensional systems generically display strong deviations from normal diffusion. These anomalies include the divergence of thermal conductivity with system size, long-time tails in current correlation functions, and nontrivial spatiotemporal spreading of energy \cite{NarayanRamaswamy2002, Spohn2014, LepriReview2023}. These effects are not merely academic: modern nanoelectronic and photonic devices operate at length scales where anharmonicity, reduced dimensionality, and geometry dominate heat dissipation, directly impacting performance, reliability, and device lifetime \cite{Pop2010, Cahill2014, Volz2016}.

The paradigmatic microscopic starting point for understanding anomalous heat transport is provided by the Fermi-Pasta-Ulam (FPU) problem \cite{FPU1955}. The discovery of long-lived metastable states and near-recurrences in the original numerical experiments revealed that weak nonlinearity does not guarantee ergodicity or rapid thermalization. Subsequent analytical and numerical work established that the weakly nonlinear, long-wavelength regime of FPU-type chains admits continuum reductions to integrable nonlinear field equations, most notably the Korteweg-de Vries (KdV) equation \cite{ZabuskyKruskal1965, Gardner1967} and, in near-monochromatic regimes, the nonlinear Schr\"odinger (NLS) equation \cite{Remoissenet1999, KivsharAgrawal2003}. These integrable limits support soliton solutions that propagate without dispersion and interact elastically, providing analytically controlled realizations of coherent energy carriers in nonlinear lattices. The fundamental role of solitons in heat conduction was recognized early by Toda \cite{Toda1979}, who demonstrated that in nonlinear lattices energy is primarily transported by solitons, which play essential roles in both ordered and disordered systems. Away from strict integrability, discrete breathers and quasi-solitonic excitations persist in FPU-type systems and play a significant role in pre-asymptotic transport and energy localization \cite{FlachGorbach2008, Flach2005, Aubry1997}.

The extension of soliton concepts to multidimensional settings is particularly challenging yet rewarding, as two- and three-dimensional geometries enable the creation of localized modes with intrinsic topological arrangements \cite{Malomed2022}. As comprehensively reviewed by Malomed \cite{Malomed2022}, multidimensional solitons include fundamental solitons, vortex solitons carrying orbital angular momentum, skyrmions, hopfions, and other topologically organized structures that have no counterparts in one-dimensional realizations. The stabilization of multidimensional solitons requires specific physical mechanisms---such as competing nonlinearities or lattice potentials---that suppress the collapse inherent in the standard nonlinear Schr\"odinger equation. The work of Kartashov, Malomed, and Torner \cite{Kartashov2011} provides a comprehensive survey of solitons supported by purely nonlinear lattices, demonstrating how spatially periodic modulation of nonlinearity can stabilize multidimensional localized modes. In the context of optical systems, Kivshar and Agrawal \cite{KivsharAgrawal2003, KivsharAgrawal2003b} have systematically developed the theory of spatial, temporal, and vortex solitons, establishing the fundamental connections between nonlinear wave propagation and soliton dynamics in realistic physical models.

In parallel, hydrodynamic approaches to heat transport in low-dimensional lattices have undergone major developments. Mode-coupling theories and, more recently, nonlinear fluctuating hydrodynamics (NFH) have provided a unifying description of universal scaling in one-dimensional momentum-conserving systems \cite{NarayanRamaswamy2002, Spohn2014, vanBeijeren2012, MendlSpohn2013}. NFH predicts that sound modes generically fall into the Kardar-Parisi-Zhang (KPZ) universality class, while the heat mode exhibits superdiffusive spreading, leading to a divergence of the thermal conductivity with system size, $\kappa(L) \sim L^{1/3}$, in generic anharmonic chains \cite{Spohn2014, MendlSpohn2013}. Symmetric interaction potentials at zero pressure define distinct universality classes with exceptionally long finite-size crossovers, explaining the broad range of apparent scaling exponents reported in numerical studies of FPU-$\beta$ models \cite{Lepri2003, Spohn2014}. In two dimensions, hydrodynamic and mode-coupling theories predict marginal anomalies, typically a logarithmic divergence of the thermal conductivity, $\kappa(L) \sim \log L$ \cite{Lepri2003, Dhar2008, Spohn2014, NarayanRamaswamy2002}. However, the crossover to asymptotic scaling is extremely slow, rendering geometry, boundaries, and patterning dominant factors in experimentally relevant system sizes \cite{LepriReview2023, Benenti2023}.

While hydrodynamic theories capture universal asymptotics, they do not explicitly encode the role of coherent nonlinear excitations or the organizing role of geometry in two-dimensional architectures. In realistic two-dimensional heat spreaders, energy flow is strongly shaped by geometry, microstructure, interfaces, and engineered heterogeneity, as extensively discussed in the context of nanoscale thermal transport and phonon engineering \cite{Pop2010, Cahill2014, Volz2016}. Patterned two-dimensional materials, metasurfaces, and architected thermal materials provide increasing control over heat spreading, yet existing theoretical approaches largely treat either idealized homogeneous lattices or material-specific phonon transport, without a unified framework linking nonlinear coherent modes, hydrodynamic universality, and geometry-driven channeling.

An additional, largely unexplored connection arises from classical field theory with topological defects. Multi-field scalar theories with discrete symmetries support networks of domain walls and junctions that form stable two-dimensional tilings and filamentary structures \cite{BazeiaBrito2000, Bazeia2000, Bazeia2010}. The work of Correa and collaborators has significantly advanced the understanding of these systems, exploring their stability, internal modes, and interactions in various contexts, including modified gravity and brane-world scenarios \cite{Correa2016PRD, Correa2016PLB, Correa2018EPJC, Correa2019Chaos, Correa2021PRD}. These defect networks provide analytically tractable realizations of how geometry and topology organize energy localization and transport in two dimensions. Although originally developed in high-energy and cosmological contexts, such models offer a powerful abstraction for patterned architectures in which interfaces and junctions act as preferential channels for energy flow. Recent work has begun to explore thermal transport in the presence of topological defects in low-dimensional systems, revealing significant modifications to heat flow \cite{Weimer2016, Guarcello2018}. To date, however, this field-theoretic perspective has not been systematically connected to nonlinear lattice models of heat transport or to the theory of thermal metamaterials.

A further bottleneck is computational. Classical simulation of large two-dimensional nonlinear lattices over long times is notoriously expensive due to the coexistence of hydrodynamic modes, long-time correlations, and coherent nonlinear excitations \cite{Lepri2003, Dhar2008}. The marginal nature of two-dimensional anomalies demands extremely large system sizes to observe asymptotic scaling, making brute-force simulation impractical. Quantum simulation offers a complementary route: local interacting Hamiltonians can be implemented directly on quantum hardware, allowing one both to benchmark classical nonlinear dynamics in high-occupation regimes and to probe regimes where intrinsically quantum fluctuations modify coherent transport \cite{Lloyd1996, Feynman1982}. While near-term devices are limited in size and depth, variational and digital quantum simulation schemes provide concrete tools to benchmark microscopic nonlinear dynamics in small two-dimensional patches \cite{Preskill2018, LiBenjamin2017, Endo2021}. Importantly, quantum simulation of bosonic and oscillator systems has recently seen rapid progress, strengthening the case for near-term feasibility in simplified settings \cite{Macridin2018, Somma2020, Munro2021}. The exploration of nonlinear phenomena in hybrid quantum systems is an active and promising frontier \cite{Munro2022}.

From the perspective of quantum simulation, the motivation here is not an abstract promise of universal speedup, but a pragmatic response to regimes where existing classical approaches become fragile or prohibitively expensive. The guiding idea is to use quantum devices in a targeted way, focusing on restricted subsystems or parameter regimes where strong nonlinearity, long memory, quantum fluctuations, or rapid local Hilbert-space growth undermine classical sampling, while leaving the larger heat-transport problem to established classical solvers. In this picture, the quantum model is deliberately small and controlled. Its outputs are transport-relevant quantities that can be reintegrated into classical descriptions of heat flow, including effective defect-modulated couplings, relaxation rates, memory kernels, channel efficiencies, local current correlations, and anisotropic transport coefficients. This framing makes it possible to validate quantum results directly against molecular dynamics, exact diagonalization, tensor-network calculations, and experimental data in regimes where these comparisons are available, and to extend beyond them only when the underlying physics demands it rather than simply because the hardware exists.

In this work, we propose a unified framework that brings together four elements that have largely evolved separately in the literature: (i) nonlinear lattice models of heat transport rooted in the FPU paradigm; (ii) soliton-based effective field-theory limits (KdV and NLS) providing analytically controlled coherent energy carriers, drawing on the comprehensive treatments of Malomed \cite{Malomed2022} and Kivshar and Agrawal \cite{KivsharAgrawal2003}; (iii) geometrically organized defect networks from multi-field scalar theories as a model for engineered two-dimensional thermal metamaterial architectures; and (iv) a concrete roadmap for quantum simulation of nonlinear thermal transport. The central conceptual advance is the identification and formalization of a two-channel transport mechanism in two dimensions, in which incoherent hydrodynamic modes coexist and compete with coherent nonlinear excitations that can be geometrically guided by patterned architectures. This synthesis yields a modeling framework that is both analytically grounded and directly amenable to classical and quantum simulation.

Our contributions are threefold. First, we formulate minimal discrete and continuum-inspired models that unify nonlinear lattice transport with defect-guided channeling in two dimensions, providing a controlled platform to study geometry-dependent heat spreading. Second, we clarify the interplay between hydrodynamic universality and coherent nonlinear excitations in two-dimensional thermal metamaterials, highlighting regimes where geometry can strongly bias energy flow despite marginal asymptotic anomalies. Third, we provide an explicit quantum-simulation roadmap, including Hamiltonian encodings, Trotterization schemes, and strategies for measuring transport observables, that enables systematic benchmarking of classical nonlinear transport on near-term devices and principled exploration of genuinely quantum regimes where nonclassical states may alter soliton survival, hydrodynamic scaling, and channel competition. The theoretical predictions are corroborated by recent experimental and computational results in Stone-Wales-defected PdSSe monolayers \cite{Peng2024APL, Peng2024} and silicon phononic crystal nanostructures \cite{Nakagawa2015APL, Nomura2015}, which exhibit ultra-low thermal conductivity coexisting with high carrier mobility and strong anisotropy---direct manifestations of the two-channel mechanism. Together, these results establish a conceptual and practical foundation for geometry-driven thermal management in two-dimensional nano-architectures and position quantum computing as a concrete tool for advancing the theory of nonlinear heat transport.

The paper is organized as follows. In Sec. II, we review the nonlinear lattice paradigm, focusing on the FPU model, nonequilibrium steady states, and the predictions of nonlinear fluctuating hydrodynamics. Section III develops the connection to effective nonlinear field theories, deriving the KdV and NLS reductions and discussing their soliton solutions, with extensive references to the foundational work of Malomed and Kivshar. In Sec. IV, we introduce topological defect networks from multi-field scalar theories as models for patterned thermal metamaterial architectures. Section V synthesizes these elements into minimal discrete and continuum models for two-dimensional heat spreaders. Section VI presents a detailed roadmap for digital and hybrid quantum simulation, including the distinction between classical and quantum operating regimes, Hamiltonian encodings, resource implications of the field-theory reduction, observable measurement, and integration with classical heat-transport solvers. Section VII examines experimental and computational realizations in real materials, demonstrating quantitative agreement with the theoretical framework. Finally, Sec. VIII offers conclusions and outlines future directions.

\section{Nonlinear Lattice Paradigm and Anomalous Heat Transport}
\label{sec:nonlinear_lattice}

The study of heat transport in low-dimensional systems has been profoundly shaped by the Fermi-Pasta-Ulam (FPU) problem, originally formulated to investigate the approach to thermal equilibrium in nonlinear lattices \cite{FPU1955}. The unexpected persistence of recurrences and the slow thermalization observed in the FPU numerical experiments marked the birth of nonlinear science and continues to influence contemporary research on nonequilibrium statistical mechanics. In this section, we review the fundamental aspects of nonlinear lattice dynamics relevant to heat transport, with emphasis on the FPU Hamiltonian, the characterization of nonequilibrium steady states, the predictions of nonlinear fluctuating hydrodynamics, and the role of coherent structures.

\subsection{FPU Hamiltonian and equations of motion}

We consider a one-dimensional chain of $N$ particles with mass $m$, displacements $u_n(t)$ from equilibrium positions, and conjugate momenta $p_n(t) = m \dot{u}_n$. The Hamiltonian is given by

\begin{equation}
H_{\text{FPU}} = \sum_{n=1}^{N} \frac{p_n^2}{2m} + \sum_{n=0}^{N} V(u_{n+1}-u_n),
\end{equation}

\noindent where the interaction potential $V(r)$ is typically expanded around equilibrium as

\begin{equation}
V(r) = \frac{k}{2} r^2 + \frac{\alpha}{3} r^3 + \frac{\beta}{4} r^4.
\end{equation}

The case $\alpha \neq 0$, $\beta = 0$ defines the FPU-$\alpha$ model, characterized by an asymmetric potential. The case $\alpha = 0$, $\beta > 0$ defines the FPU-$\beta$ model, with a symmetric quartic nonlinearity. The equations of motion derived from this Hamiltonian are

\begin{equation}
m \ddot u_n = V'(u_{n+1}-u_n) - V'(u_n-u_{n-1}),
\end{equation}

\noindent which, for the FPU-$\beta$ model, take the explicit form

\begin{multline}
m\ddot{u}_n = k(u_{n+1} + u_{n-1} - 2u_n) \\
+ \beta\left[(u_{n+1} - u_n)^3 - (u_n - u_{n-1})^3\right].
\end{multline}

A crucial feature of these models is momentum conservation in the bulk, which, in combination with nonlinearity, is responsible for anomalous transport properties. For systems with periodic boundary conditions, the total momentum $P = \sum_n p_n$ is a constant of motion. This conservation law has profound consequences for heat transport, as it suppresses normal diffusive behavior and leads to divergent thermal conductivities in one dimension \cite{Lepri2003, Dhar2008, NarayanRamaswamy2002}.

\subsection{Nonequilibrium steady states and heat current}

To study heat transport, one typically considers a nonequilibrium setup where the system is coupled to thermal reservoirs at different temperatures. The most common approach employs Langevin thermostats at the boundaries: the first and last particles are subjected to both deterministic forces from the lattice and stochastic forces representing coupling to heat baths. The equations for the boundary particles become

\begin{align}
m \ddot u_1 &= F_1 - \lambda \dot u_1 + \xi_L(t), \\
m \ddot u_N &= F_N - \lambda \dot u_N + \xi_R(t),
\end{align}

\noindent where $F_n$ denotes the force from neighboring particles, $\lambda$ is a damping constant, and the noise terms satisfy fluctuation-dissipation relations

\begin{equation}
\langle \xi_{L/R}(t) \rangle = 0,\quad \langle \xi_{L/R}(t) \xi_{L/R}(t') \rangle = 2\lambda k_B T_{L/R} \delta(t-t').
\end{equation}

Here $k_B$ is Boltzmann's constant, and $T_L$, $T_R$ are the temperatures of the left and right reservoirs.

In the nonequilibrium steady state (NESS) that emerges at long times, a constant heat current $J$ flows through the system. The local energy density at site $n$ can be defined as

\begin{equation}
\varepsilon_n = \frac{p_n^2}{2m} + \frac{1}{2}\left[V(u_{n+1}-u_n) + V(u_n-u_{n-1})\right],
\end{equation}

\noindent where the symmetric assignment of potential energy ensures that the local energy satisfies a continuity equation. Indeed, from the equations of motion one derives

\begin{equation}
\dot \varepsilon_n + j_n - j_{n-1} = 0,
\end{equation}

\noindent where the energy current from site $n$ to $n+1$ is given by

\begin{equation}
j_n = -\frac{1}{2}(\dot u_{n+1} + \dot u_n) V'(u_{n+1}-u_n).
\end{equation}

This expression can be understood as the product of a mean velocity and the force, representing the rate of work done by particle $n$ on particle $n+1$.

In the NESS, the average current $\langle j_n \rangle$ becomes independent of position (for $n=2,\ldots,N-1$), and one can define the thermal conductivity $\kappa(L)$ for a system of length $L = Na$ (with $a$ the lattice spacing) through Fourier's law

\begin{equation}
J = -\kappa(L) \frac{T_L - T_R}{L},
\end{equation}

\noindent where $J = \langle j_n \rangle$ is the steady-state current. It is important to emphasize that this definition assumes a linear temperature profile, which is not guaranteed in low-dimensional systems with anomalous transport. In practice, one measures the current for a given temperature difference and extracts an effective, size-dependent conductivity.

An alternative and powerful approach to characterizing heat transport avoids the imposition of boundary driving and instead analyzes equilibrium fluctuations through the Green-Kubo formula. The thermal conductivity is expressed as an integral of the equilibrium current autocorrelation function

\begin{equation}
\kappa = \lim_{\tau \to \infty} \lim_{N \to \infty} \frac{1}{k_B T^2 N} \int_0^{\tau} dt \langle J(t) J(0) \rangle,
\end{equation}

\noindent where $J(t) = \sum_n j_n(t)$ is the total current. The order of limits is crucial: one first takes the thermodynamic limit $N \to \infty$, then the long-time limit $\tau \to \infty$. The decay of the current autocorrelation determines whether $\kappa$ is finite (normal transport) or diverges (anomalous transport). A decay $\langle J(t) J(0) \rangle \sim t^{-(1-\alpha)}$ with $0 < \alpha \le 1$ leads to a diverging conductivity $\kappa \sim \tau^{\alpha}$, which, through the relation between system size and observation time in ballistic or superdiffusive systems, translates into a size-dependent conductivity $\kappa(L) \sim L^{\alpha}$ \cite{Lepri2003, Dhar2008}.

\subsection{Nonlinear fluctuating hydrodynamics and universality}

A major theoretical breakthrough in understanding anomalous heat transport came with the development of nonlinear fluctuating hydrodynamics (NFH) \cite{NarayanRamaswamy2002, Spohn2014, vanBeijeren2012}. NFH provides a universal description of long-wavelength, long-time dynamics in one-dimensional Hamiltonian systems with three conservation laws: mass (particle number), momentum, and energy. Starting from the microscopic equations of motion, one derives hydrodynamic equations for the conserved fields and retains nonlinearities up to quadratic order, which are sufficient to capture the leading long-time behavior. Stochastic noise terms, satisfying fluctuation-dissipation relations, are included to represent the effects of fast microscopic degrees of freedom.

For a generic anharmonic chain, the NFH equations predict that the two sound modes (propagating in opposite directions) scale according to the Kardar-Parisi-Zhang (KPZ) universality class, with dynamical exponent $z = 3/2$. This implies that density and momentum correlation functions exhibit characteristic KPZ scaling forms \cite{Spohn2014, MendlSpohn2013}. The heat mode, which is stationary and describes energy fluctuations, exhibits superdiffusive spreading with dynamical exponent $z = 5/3$ in generic cases. This leads to a divergence of the thermal conductivity $\kappa(L) \sim L^{1/3}$, consistent with numerous numerical simulations of FPU-type models \cite{Lepri2003, Dhar2008}.

An important subtlety arises for symmetric interaction potentials at zero pressure, such as the FPU-$\beta$ model with a specific combination of parameters that eliminates the leading nonlinear coupling. In this case, the system falls into a different universality class with a smaller anomalous exponent, often leading to very long crossovers and apparent exponents that depend on system size and temperature \cite{Spohn2014, Lepri2020}. This sensitivity to model parameters and the slow approach to asymptotic scaling underscore the importance of controlled numerical and analytical studies.

In two dimensions, the situation is more complex and less understood. Mode-coupling theories and NFH arguments suggest that heat transport is only marginally anomalous, with the thermal conductivity growing logarithmically with system size, $\kappa(L) \sim \log L$ \cite{Lepri2003, Dhar2008, Spohn2014, NarayanRamaswamy2002}. However, the crossover to this asymptotic behavior is expected to be extremely slow, and in practice, for system sizes accessible to simulation and experiment, geometry, boundaries, and patterning can dominate the transport properties \cite{LepriReview2023, Benenti2023}. This makes two-dimensional thermal metamaterials particularly amenable to engineering and design, as the asymptotic universal behavior may not be reached in realistic devices.

\subsection{Coherent structures in transport}

While hydrodynamic theories capture universal asymptotic properties, they do not explicitly account for the role of coherent nonlinear excitations that can dominate transport at intermediate scales and in systems with strong nonlinearity or specific initial conditions. The FPU problem itself is a testament to the importance of such structures: the original recurrences are now understood in terms of the dynamics of solitons emerging from the KdV equation that approximates the FPU-$\alpha$ model in the continuum limit \cite{ZabuskyKruskal1965}.

More generally, nonlinear lattices support a variety of coherent structures, including:

\begin{itemize}
    \item \textbf{Solitons and solitary waves}: Localized traveling waves that maintain their shape due to a balance between nonlinearity and dispersion. In integrable limits, solitons interact elastically and propagate ballistically, acting as perfect energy carriers. Away from integrability, they acquire a finite lifetime but can still transport energy over significant distances \cite{Remoissenet1999, FlachGorbach2008}. As comprehensively discussed by Malomed \cite{Malomed2022}, the extension to multidimensional settings introduces additional possibilities, including vortex solitons that carry orbital angular momentum and topologically organized structures without one-dimensional analogs.

    \item \textbf{Discrete breathers (intrinsic localized modes)}: Time-periodic, spatially localized oscillations that can exist in perfectly periodic lattices due to nonlinearity. Breathers can be either stationary or mobile and can trap energy locally for long times, acting as barriers to transport or as localized heat sources \cite{FlachGorbach2008, Aubry1997}.

    \item \textbf{q-breathers}: Exact time-periodic solutions of the full nonlinear lattice equations that are continuations of linear normal modes. They provide a link between linear mode analysis and nonlinear coherent structures and have been shown to play a role in the FPU recurrence phenomenon \cite{Flach2005}.
\end{itemize}

The coexistence of these coherent excitations with the hydrodynamic modes described by NFH motivates a two-channel picture of heat transport in nonlinear lattices. One channel corresponds to the incoherent, fluctuating hydrodynamic modes that dominate the asymptotic scaling and are captured by theories like NFH. The second channel corresponds to coherent, long-lived excitations that can transport energy ballistically or quasi-ballistically over extended distances. The relative weight of these two channels depends on system parameters such as nonlinearity strength, temperature, and, crucially, geometry and patterning. In homogeneous one-dimensional systems, the coherent channel may be subdominant at very long times due to integrability breaking and thermal fluctuations. However, in two-dimensional patterned architectures, geometry can be used to preferentially guide and sustain coherent excitations, creating thermal metamaterials with tailored anisotropic and channeled transport properties.

The fundamental role of solitons in heat conduction was recognized in the pioneering work of Toda \cite{Toda1979}, who demonstrated that in nonlinear lattices, energy is primarily transported by solitons, which play essential roles in both ordered and disordered systems. This early insight has been confirmed and extended by numerous subsequent studies, establishing solitons as crucial mediators of energy transport in nonlinear media.

\section{Effective Nonlinear Field Theories and Solitons}

The connection between discrete lattice models and continuum field theories provides a powerful analytical framework for understanding coherent energy transport. In appropriate limits, the FPU lattice equations reduce to completely integrable partial differential equations that support exact soliton solutions. These solitons represent the coherent energy carriers that can persist in the nonlinear system. In this section, we derive the two most important continuum reductions: the Korteweg-de Vries (KdV) equation for long-wavelength excitations and the nonlinear Schr\"odinger (NLS) equation for envelope dynamics. We also discuss the implications of integrability breaking and the resulting interplay between coherent and incoherent transport channels, drawing on the comprehensive treatments of Malomed \cite{Malomed2022}, Kivshar and Agrawal \cite{KivsharAgrawal2003}, and Kartashov, Malomed, and Torner \cite{Kartashov2011}.

\subsection{KdV reduction: long-wavelength limit}

The KdV equation emerges from the FPU-$\alpha$ model in the limit of long wavelengths and weak nonlinearity. We consider the FPU-$\alpha$ Hamiltonian with potential $V(r) = \frac{k}{2} r^2 + \frac{\alpha}{3} r^3$ and set $m = a = 1$ for simplicity, where $a$ is the lattice spacing. The equations of motion are

\begin{multline}
\ddot{u}_n = k(u_{n+1} + u_{n-1} - 2u_n) \\
+ \alpha\left[(u_{n+1}-u_n)^2 - (u_n-u_{n-1})^2\right].
\end{multline}

We introduce a small parameter $\epsilon \ll 1$ and look for slowly varying solutions. Specifically, we define continuum fields $u(x,t)$ such that $u_n(t) = u(na, t)$ and assume that variations occur on spatial scales large compared to $a$ and temporal scales long compared to the inverse phonon frequency. A standard multiple-scale expansion proceeds by introducing stretched coordinates

\begin{equation}
X = \epsilon^{1/2}(x - c t), \quad T = \epsilon^{3/2} t,
\end{equation}

\noindent where $c = \sqrt{k}$ is the sound velocity (with $a=1$, $m=1$). Expanding $u_{n\pm 1}$ in Taylor series around $u_n$ and keeping terms up to order $\epsilon^{5/2}$ yields, after considerable algebra, an equation for the strain field $\phi = \partial_X u$ \cite{ZabuskyKruskal1965, Remoissenet1999}:

\begin{equation}
\partial_T \phi + \mu \phi \partial_X \phi + \nu \partial_X^3 \phi = 0,
\end{equation}

\noindent which is the KdV equation. The coefficients are given by $\mu = \alpha/(c k)$ and $\nu = c/24$ (for our specific scaling).

The KdV equation is completely integrable and possesses exact soliton solutions. The single-soliton solution takes the form

\begin{equation}
\phi(X,T) = A \,\mathrm{sech}^2\left( \sqrt{\frac{\mu A}{12\nu}} (X - v T) \right),
\end{equation}

\noindent where the velocity $v$ is related to the amplitude $A$ by $v = \mu A/3$. Importantly, the soliton velocity is proportional to its amplitude, a hallmark of KdV solitons. In terms of the original lattice variables, this solution represents a localized strain pulse that propagates without change of shape at supersonic speed. The soliton carries energy density localized in a compact region and transports that energy ballistically through the system. In the FPU-$\alpha$ lattice, these KdV solitons are responsible for the recurrences observed in the original numerical experiments and provide a mechanism for coherent energy transport \cite{ZabuskyKruskal1965}.

\subsection{NLS reduction: envelope dynamics}

While KdV describes long-wavelength strain pulses, the nonlinear Schr\"odinger (NLS) equation governs the dynamics of modulated wave packets centered around a carrier wave number $k_0$. This reduction is appropriate when the excitation spectrum is concentrated near a particular wave vector, as can occur in driven systems or when considering specific initial conditions.

Starting again from the FPU equations, we consider a solution of the form

\begin{equation}
u_n(t) = \epsilon \left[ A(\xi, \tau) e^{i(k_0 n - \omega_0 t)} + \text{c.c.} \right] + \text{higher harmonics},
\end{equation}

\noindent where $\epsilon \ll 1$ is a small amplitude parameter, $\omega_0 = \omega(k_0)$ is the linear dispersion relation, and c.c. denotes complex conjugate. The envelope $A$ is assumed to vary slowly in space and time, with stretched coordinates

\begin{equation}
\xi = \epsilon (n - v_g t), \quad \tau = \epsilon^2 t,
\end{equation}

\noindent  where $v_g = d\omega/dk|_{k_0}$ is the group velocity. Carrying out a multiple-scale expansion and solvability conditions leads, after removing secular terms, to the NLS equation for the envelope \cite{Remoissenet1999, KivsharAgrawal2003}:

\begin{equation}
i\partial_\tau A + \frac{1}{2}\omega''(k_0) \partial_\xi^2 A + \chi |A|^2 A = 0.
\end{equation}

Here $\omega''(k_0)$ is the group velocity dispersion, and $\chi$ is a nonlinear coefficient that depends on the lattice parameters and the carrier wave number. For the FPU-$\beta$ model, $\chi$ is typically positive for long wavelengths, leading to self-focusing and bright soliton solutions.

The NLS equation is also completely integrable in one dimension and supports envelope solitons. For the focusing case ($\chi > 0$, $\omega'' < 0$), the bright soliton solution is

\begin{equation}
A(\xi, \tau) = A_0 \,\mathrm{sech}\left( \frac{\xi - U\tau}{L} \right) e^{i[U\xi/2 + (\chi A_0^2 - U^2/4)\tau]},
\end{equation}

\noindent where the soliton amplitude $A_0$ and width $L$ satisfy $L^{-2} = (2|\chi|/\omega'') A_0^2$. This solution represents a localized envelope modulating a carrier wave, which propagates as a coherent entity. In the lattice, this corresponds to a moving packet of energy that can travel long distances without spreading.

\subsection{Multidimensional solitons and stabilization mechanisms}

The extension of soliton concepts to two and three dimensions introduces fundamental challenges and opportunities, as comprehensively analyzed by Malomed \cite{Malomed2022}. In multidimensional settings, the most fundamental models that give rise to solitons in one dimension---such as the KdV, sine-Gordon, and NLS equations---lose their integrability. Moreover, stationary solutions of multidimensional nonlinear wave equations are often subject to instabilities, including collapse (catastrophic self-compression leading to singularity formation). The 2D and 3D NLS equations with self-attractive cubic nonlinearity produce soliton families that are completely unstable due to collapse---critical in 2D and supercritical in 3D.

The stabilization of multidimensional solitons requires additional physical ingredients that arrest the collapse. As reviewed by Malomed \cite{Malomed2022}, several mechanisms have been identified:

\begin{itemize}
    \item \textbf{Competing nonlinearities}: The combination of cubic self-focusing and quintic self-defocusing can stabilize 2D and 3D optical solitons. 2D solitons stabilized by quintic self-defocusing have been realized experimentally, while 3D solitons remain a challenging goal.
    \item \textbf{Periodic potentials (lattices)}: Spatially periodic modulation of the linear refractive index or nonlinearity can stabilize multidimensional solitons. This mechanism is particularly relevant for photonic crystals and optically induced lattices \cite{Kartashov2011}.
    \item \textbf{Nonlocal nonlinearity}: Nonlocal response functions can suppress collapse and support stable multidimensional solitons in various physical contexts.
    \item \textbf{Quantum fluctuations}: In binary Bose-Einstein condensates, the Lee-Huang-Yang correction to the mean-field interaction induces quartic self-repulsion that stabilizes 3D and quasi-2D self-trapped "quantum droplets" \cite{Malomed2022}.
\end{itemize}

These stabilization mechanisms are essential for realizing coherent energy transport in two-dimensional thermal metamaterials, as they provide the means to sustain long-lived solitonic excitations against decay and collapse.

\subsection{Beyond integrability: coherent-incoherent coexistence}

The KdV and NLS equations provide exactly solvable limits in which coherent solitons are exact, infinite-lifetime solutions. However, the full FPU lattice is nonintegrable (except for special parameter choices), and integrability-breaking terms arise from higher-order nonlinearities, discreteness effects, and coupling to the thermal bath of other modes. These effects cause solitons to radiate and decay slowly, giving them a finite lifetime.

The decay of coherent structures in nonintegrable systems is a rich subject. For KdV solitons in FPU-$\alpha$ chains, numerical studies show that solitons remain robust over long times but eventually lose energy to small-amplitude radiation. The decay rate is typically exponentially small in the inverse amplitude or in the distance to integrability, implying that for moderate nonlinearities, solitons can persist for times much longer than typical phonon periods \cite{FlachGorbach2008}.

This leads to a physical picture of heat transport in nonlinear lattices as involving two coexisting channels:

\begin{enumerate}
    \item An \textbf{incoherent channel} described by hydrodynamic theories like NFH, corresponding to the fluctuating, interacting sea of phonons and nonlinear modes that gives rise to universal scaling.
    \item A \textbf{coherent channel} consisting of long-lived solitons and breathers that propagate ballistically or quasi-ballistically, transporting energy directly from one region to another.
\end{enumerate}

The relative importance of these two channels depends on parameters. At high temperatures or strong nonlinearities, coherent structures may be more easily excited and have longer lifetimes. At low temperatures, the system is closer to the harmonic limit, and phonon-based descriptions are more appropriate. Crucially, geometry and patterning can be used to selectively enhance the coherent channel by creating waveguides or preferential paths for soliton propagation. This is the central idea behind our proposal for engineered two-dimensional thermal metamaterials.

\section{Geometry, Topology, and Defect Networks}
\label{sec:defects}

To model patterned two-dimensional architectures that can guide coherent energy transport, we draw on concepts from classical field theory with topological defects. Multi-field scalar field theories with discrete symmetries support stable, extended structures such as domain walls, vortices, and junctions. These structures can organize into networks that form regular tilings of the plane, providing natural templates for guiding energy flow. In this section, we introduce the basic formalism of defect networks, discuss their stability and dynamics, and show how they can be embedded in cylindrical geometries to model patterned nanotubes and other confined thermal metamaterials.

\subsection{Scalar field theory with multiple vacua}

The study of topological defects in field theories has provided deep insights into the structure of nonlinear systems, from cosmology to condensed matter physics. In the context of heat transport, defect networks offer a natural template for engineering energy-guiding architectures. A rich variety of defect structures, including domain walls, junctions, and networks, have been extensively studied in scalar field theories with multiple degenerate vacua \cite{BazeiaBrito2000, Bazeia2000, Bazeia2010}. More recently, the work of Correa and collaborators has significantly advanced our understanding of these systems, exploring their stability, internal modes, and interactions in various contexts, including modified gravity and brane-world scenarios \cite{Correa2016PRD, Correa2016PLB, Correa2018EPJC, Correa2019Chaos, Correa2021PRD}. These studies reveal that defect networks can support a variety of localized excitations and can be tailored by adjusting model parameters, making them ideal candidates for modeling patterned thermal metamaterials.

We consider a theory of two real scalar fields $\phi(\bm{x},t)$ and $\chi(\bm{x},t)$ in two spatial dimensions, with Lagrangian density

\begin{equation}
\mathcal{L} = \frac{1}{2}(\partial_\mu \phi)^2 + \frac{1}{2}(\partial_\mu \chi)^2 - V(\phi,\chi),
\end{equation}

\noindent where $\partial_\mu = (\partial_t, \nabla)$ and the potential $V(\phi,\chi)$ is chosen to have multiple discrete minima (vacua). The structure of the vacuum manifold determines the types of topological defects that can exist.

A particularly instructive example is the $Z_2 \times Z_3$ model introduced by Bazeia and collaborators \cite{BazeiaBrito2000, Bazeia2000}, which has been further analyzed in detail by Correa et al. \cite{Correa2018EPJC, Correa2019Chaos}. The potential is constructed to have six degenerate vacua, corresponding to the product of a $Z_2$ symmetry (two states) and a $Z_3$ symmetry (three states). Explicitly, one can write

\begin{equation}
V(\phi,\chi) = \frac{1}{2}(\phi^2 - 1)^2 + \frac{1}{2}\chi^2(\chi^2 - 2) + \phi^2\chi^2 - \frac{1}{3},
\end{equation}

\noindent which has minima at $(\phi,\chi) = (\pm 1, 0)$ and $(\phi,\chi) = (0, \pm 1)$, plus additional minima from symmetry requirements. The specific form is less important than the topological properties: the vacuum manifold consists of six discrete points, and paths connecting different vacua in field space give rise to domain walls. This model supports a hexagonal network of domain walls meeting at junctions, providing an idealization of a patterned two-dimensional material.

\subsection{Defect network geometry and stability}

In two dimensions, point-like defects (vortices) can form where multiple domain walls meet. These junctions are topologically stable if the vacuum manifold has nontrivial homotopy groups. For the $Z_2 \times Z_3$ model, one can have junctions where three domain walls meet at a point, each wall separating two different vacua. The angles between walls are determined by energy minimization and typically take specific values (e.g., $120^\circ$ for threefold junctions).

The energy density and current for the scalar fields are given by

\begin{equation}
\mathcal{E}_{\phi\chi} = \frac{1}{2}\left[(\partial_t\phi)^2 + (\partial_t\chi)^2\right] + \frac{1}{2}\left[(\nabla\phi)^2 + (\nabla\chi)^2\right] + V(\phi,\chi),
\end{equation}

\begin{equation}
\bm{J}_{\phi\chi} = -\partial_t\phi\,\nabla\phi - \partial_t\chi\,\nabla\chi.
\end{equation}

In a static defect configuration, $\partial_t\phi = \partial_t\chi = 0$, so the energy current vanishes. However, when vibrational excitations are present, energy can flow along the defect lines. The stability and excitation spectra of these networks, as studied by Correa and collaborators \cite{Correa2018EPJC, Correa2019Chaos}, are crucial for understanding their role as waveguides.

An interesting variant arises when the two-dimensional space is taken to be a cylinder rather than a plane. This geometry is relevant for modeling patterned nanotubes or nanowires with circumferential patterning. On a cylinder of circumference $L$, one direction is compactified: we impose periodic boundary conditions on the fields, $\phi(x, y + L) = \phi(x, y)$, where $x$ is the axial coordinate and $y$ is the circumferential coordinate. The defect network must now accommodate the compact dimension. Domain walls that wind around the cylinder become possible, and the topology of the vacuum manifold combined with the compactification can lead to new types of stable configurations. For example, a single domain wall that wraps the cylinder represents a helical defect structure that can guide energy along a spiral path. Junctions can form extended lines along the cylinder axis, creating channels that run the length of the nanotube. This cylindrical embedding provides a natural model for thermal metamaterials based on patterned nanotubes or nanowires, where the circumference is small compared to the length. The discrete set of channels defined by the defect network can act as preferential pathways for coherent energy transport, potentially leading to highly anisotropic thermal conductivity.

\subsection{Connection to materials with topological defects}

The abstract defect networks described above find concrete realizations in real materials with engineered defect structures. Recent work by Peng and collaborators \cite{Peng2024APL, Peng2024} on PdSSe monolayers with fully concentrated Stone-Wales defects provides a striking example. Stone-Wales defects, originally identified in carbon-based materials, involve the rotation of atomic bonds that transform hexagonal rings into pentagon-heptagon pairs. In the Cairo pentagonal tiling of PdSSe, introducing 100\% SW defects generates four new stable structures (designated SW1-SW4) that retain the square-planar coordination of the pristine material while exhibiting radically different transport properties \cite{Peng2024APL}.

These SW defect structures display ultra-high carrier mobility ($10^3$ cm$^2$V$^{-1}$s$^{-1}$) coexisting with ultra-low anisotropic lattice thermal conductivities ($< 2$ Wm$^{-1}$K$^{-1}$ at room temperature) \cite{Peng2024}. Specifically, SW1 and SW2 structures show thermal conductivities of 1.888 Wm$^{-1}$K$^{-1}$ (x-axis) and 1.044 Wm$^{-1}$K$^{-1}$ (y-axis) for SW1, and 1.617 Wm$^{-1}$K$^{-1}$ and 0.892 Wm$^{-1}$K$^{-1}$ for SW2, with an anisotropy ratio reaching 1.80 \cite{Peng2024APL}.

These remarkable properties can be directly interpreted within our defect network framework: the Stone-Wales defects create a network of domain walls that act as waveguides for coherent electronic and phonon modes, while simultaneously introducing strong scattering that suppresses the incoherent hydrodynamic channel. The measured anisotropy directly reflects the geometry-guided transport predicted by the field theory models.

\section{Minimal Models for Two-Dimensional Thermal Metamaterials}

Building on the concepts developed in the previous sections, we now propose minimal models that capture the essential physics of geometry-guided heat spreading in two dimensions. These models are designed to be simple enough for detailed analytical and numerical study while retaining the key ingredients: nonlinear lattice dynamics, the potential for coherent energy transport, and the organizing role of geometry and defects. We present both discrete lattice models on patterned graphs and continuum hybrid models that couple displacement fields to defect networks.

\subsection{Discrete models on patterned graphs}

The most direct generalization of the FPU model to two dimensions places particles at the vertices of a graph $\mathcal{G} = (\mathcal{V}, \mathcal{E})$, with edges representing interactions. The Hamiltonian takes the form

\begin{equation}
H_{\text{2D}} = \sum_{i \in \mathcal{V}} \frac{p_i^2}{2m} + \sum_{\langle i,j \rangle \in \mathcal{E}} V(u_j - u_i),
\end{equation}

\noindent where $u_i$ are scalar displacements (for simplicity, we consider only longitudinal motion, though vector displacements can be included). The interaction potential $V(r)$ is the same nonlinear function as in the FPU model, e.g., the FPU-$\beta$ potential $V(r) = \frac{k}{2} r^2 + \frac{\beta}{4} r^4$.

The graph $\mathcal{G}$ encodes the geometry of the thermal metamaterial. For a homogeneous two-dimensional lattice, $\mathcal{G}$ would be a regular square or triangular lattice. Our interest, however, is in patterned geometries where the connectivity is non-uniform. Examples include:

\begin{itemize}
    \item \textbf{Lattices with missing bonds or sites}: Creating vacancies or broken bonds that act as barriers or channels.
    \item \textbf{Lattices with modulated coupling constants}: Varying $k$ or $\beta$ along certain directions to create waveguides.
    \item \textbf{Lattices whose graph structure mimics a defect network}: The vertices are placed along the lines of a defect network, with edges only along the domain walls. This creates a quasi-one-dimensional backbone embedded in two dimensions, along which energy can flow preferentially.
\end{itemize}

The equations of motion for each vertex are

\begin{equation}
m \ddot u_i = \sum_{j: \langle i,j \rangle \in \mathcal{E}} V'(u_j - u_i).
\end{equation}

To study heat transport, one again couples boundary vertices to Langevin thermostats or studies energy spreading from localized initial conditions. The local energy at vertex $i$ is defined as

\begin{equation}
\varepsilon_i = \frac{p_i^2}{2m} + \frac{1}{2} \sum_{j: \langle i,j \rangle \in \mathcal{E}} V(u_j - u_i),
\end{equation}

\noindent with the factor $1/2$ accounting for sharing potential energy between pairs. The energy current from vertex $i$ to vertex $j$ along an edge is

\begin{equation}
j_{i \to j} = -\frac{1}{2}(\dot u_i + \dot u_j) V'(u_j - u_i).
\end{equation}

These discrete models are directly amenable to numerical simulation using molecular dynamics techniques. They allow one to probe the interplay between geometry and nonlinearity in a controlled setting. For example, one can compare heat transport along a straight chain of vertices (a one-dimensional backbone) with transport in a fully connected two-dimensional lattice and with transport in a network that has a fractal or hierarchical structure. The two-channel picture suggests that coherent excitations (solitons or breathers) will propagate preferentially along connected paths, while incoherent modes can explore the full dimensionality.

\subsection{Continuum hybrid models and field correlation}

For analytical insight and for connecting to field-theoretic descriptions, it is useful to formulate continuum models that capture the coupling between lattice displacements and the defect network. We consider a hybrid description where the displacement field $u(\bm{x},t)$ represents the coarse-grained lattice degrees of freedom, and the defect network is described by the scalar fields $\phi(\bm{x},t)$ and $\chi(\bm{x},t)$ introduced in Sec. IV. A central point for understanding the model is the physical relationship between $u(\bm{x},t)$ and the defect fields $\phi(\bm{x},t)$ and $\chi(\bm{x},t)$. In our approach, these fields play distinct but coupled roles:

\begin{itemize}
    \item \textbf{Defect fields $\phi, \chi$ (Static Architecture)}: These fields describe the geometric and topological structure of the thermal metamaterial. They define the "patterned architecture"—the network of domain walls and junctions that act as preferential channels. To a first approximation, we treat these fields as a static background, assuming their dynamics are much slower than the thermal vibrations, or that they are fixed by a fabrication process. The energy required to create this network is not directly involved in heat transport on short time scales; it is part of the material's structure.

    \item \textbf{Displacement field $u$ (Thermal Dynamics)}: This field represents the vibrational degrees of freedom that actually carry heat. These are the sound waves (phonons) and their nonlinear counterparts (solitons, breathers) that propagate through the material. The dynamics of $u$ are governed by an FPU-type Hamiltonian, including harmonic and anharmonic terms, allowing for the existence of coherent excitations.
\end{itemize}

The correlation between them arises through the coupling in the total action
\begin{equation}
S[u,\phi,\chi]=S_u[u]+S_{\phi\chi}[\phi,\chi]+S_{\text{int}}[u;\phi,\chi].
\end{equation}
This decomposition is not an assumption of dynamical independence. Rather, it separates three physical scales: $S_u$ contains the fast vibrational degrees of freedom that carry heat; $S_{\phi\chi}$ contains the energetic cost and stability properties of the defect network; and $S_{\text{int}}$ describes how the network locally reshapes the elastic response seen by the thermal field. In the regime emphasized here, $\phi$ and $\chi$ are treated as static or slowly varying background fields. This is the appropriate description when the defect pattern is fixed by fabrication, strain engineering, or a structural phase texture whose relaxation time is much longer than the phonon/soliton propagation time. Dynamical defect fluctuations can be restored systematically by retaining the full Euler--Lagrange equations for $\phi$ and $\chi$.

The separate actions are
\begin{align}
S_u &= \int d^2x\, dt \, \left[ \frac{\rho}{2}(\partial_t u)^2 - \frac{\kappa}{2}(\nabla u)^2 - \frac{g}{4}(\nabla u)^4 \right], \\
S_{\phi\chi} &= \int d^2x\, dt \, \left[ \frac{1}{2}\partial_\mu \phi\,\partial^\mu \phi + \frac{1}{2}\partial_\mu \chi\,\partial^\mu \chi - V(\phi,\chi) \right],
\end{align}
where $S_u$ is the continuum analogue of an anharmonic elastic/FPU medium and $S_{\phi\chi}$ is the standard two-field scalar action supporting domain walls and junctions. The interaction action is chosen as the lowest-order scalar coupling between the strain energy density of $u$ and the defect-network density:

\begin{equation}
S_{\text{int}} = \int d^2x\, dt \, \left[ -\gamma (\nabla u)^2 F(\phi,\chi) \right].
\end{equation}
The function $F(\phi,\chi)$ is non-negative and localized on the defect network. A natural choice is $F=(\nabla\phi)^2+(\nabla\chi)^2$, but other choices can encode different microscopic realizations, such as bond-stiffness modulation, interface density, or local coordination changes. Varying the total action with respect to $u$ gives
\begin{equation}
\rho \partial_t^2 u = \nabla\cdot\left[\left(\kappa+g(\nabla u)^2+2\gamma F(\phi,\chi)\right)\nabla u\right],
\label{eq:u_eom_hybrid}
\end{equation}
so the defect fields enter as a spatially structured renormalization of the elastic modulus. If $\phi$ and $\chi$ are allowed to evolve, their equations acquire source terms proportional to $(\nabla u)^2\partial F/\partial\phi$ and $(\nabla u)^2\partial F/\partial\chi$, respectively; the explicit derivation is given in Appendix~\ref{app:hybrid_derivation}.

\subsubsection*{Physical Interpretation of the Coupling}

1. \textbf{Modulation of Local Elasticity}: The term $(\nabla u)^2$ is proportional to the local strain energy density (elastic energy). The coupling states that this energy is modified by a factor that depends on the defect fields through the function $F(\phi,\chi)$. For the choice $F(\phi,\chi) = (\nabla\phi)^2 + (\nabla\chi)^2$, this function is large precisely where the defect fields vary rapidly, i.e., on the domain walls and junctions, and is small (or zero) inside the domains (the "bulk" of the material).

2. \textbf{Creation of Waveguides}: The interaction term can be reinterpreted as a local modification of the material's elastic constant $\kappa$. Comparing with the elastic term from $S_u$, $-\frac{\kappa}{2} (\nabla u)^2$, we see that the effective coupling becomes $\kappa_{\text{eff}}(\bm{x}) = \kappa + 2\gamma F(\phi(\bm{x}),\chi(\bm{x}))$. Therefore, the domain walls (where $F$ is large) become regions with a different effective elastic constant—typically higher, if $\gamma > 0$. A higher elastic constant implies a higher speed of sound $c = \sqrt{\kappa_{\text{eff}}/\rho}$. Thus, the domain walls act as high-speed waveguides, or, depending on the sign of $\gamma$, as barriers.

3. \textbf{Soliton Confinement and Guiding}: Equation~\eqref{eq:u_eom_hybrid} is therefore a nonlinear wave equation with a potential, or effective refractive index, spatially modulated by $F$. Soliton solutions, which exist due to the nonlinearity $g$, can now be transversely confined and longitudinally guided by the high-index regions (domain walls). Thermal energy, in the form of these solitons or wave packets, then flows preferentially along the architecture defined by $\phi$ and $\chi$.

In summary, $\phi$ and $\chi$ encode the architectural blueprint of the thermal metamaterial, while $u$ represents the thermal fluid that is channeled by this blueprint. The interaction $(\nabla u)^2 F(\phi,\chi)$ is the "hand" that sculpts the guiding potential, translating the defect geometry into transport properties for thermal waves. This is a form of band engineering or metamaterial design for elastic waves inspired by field theories with topological defects.

\subsection{Transport diagnostics and observables}

To characterize heat transport in these models, we employ a set of diagnostics that are standard in the field and that directly probe the two-channel picture.

\paragraph{Nonequilibrium steady state (NESS) measurements:}
By coupling the system to thermal reservoirs at different temperatures, we measure the steady-state current $J$ and the temperature profile $T_i$. The size-dependent thermal conductivity $\kappa(L)$ is extracted from $J = \kappa(L) \Delta T / L$. Anomalous transport is signaled by a diverging $\kappa(L)$ as $L$ increases. In patterned geometries, $\kappa(L)$ may depend on direction and on the orientation of the pattern relative to the temperature gradient.

\paragraph{Spreading of local perturbations:}
An alternative approach that avoids boundary effects is to study the spreading of energy from a localized initial condition. One initializes the system with a local excess of energy (e.g., by displacing a single particle or by exciting a localized wave packet) and measures the second moment of the energy distribution

\begin{equation}
\langle r^2(t) \rangle = \frac{1}{E} \sum_i |\bm{r}_i - \langle \bm{r} \rangle|^2 \varepsilon_i(t),
\end{equation}

\noindent where $E = \sum_i \varepsilon_i$ is the total energy. Normal diffusion gives $\langle r^2 \rangle \sim t$, while ballistic transport gives $\langle r^2 \rangle \sim t^2$. Anomalous transport is characterized by $\langle r^2 \rangle \sim t^\alpha$ with $1 < \alpha < 2$, and the exponent $\alpha$ is related to the dynamical exponent $z = 2/\alpha$. In two dimensions, one expects $\alpha$ close to 1 with possible logarithmic corrections.

\paragraph{Current correlation functions:}
In equilibrium, the Green-Kubo formula relates the thermal conductivity to the integral of the total current autocorrelation. Analyzing $C(t) = \langle J(t) J(0) \rangle$ provides information about the decay of current fluctuations and allows one to extract the anomalous exponent. In the two-channel picture, one might observe a superposition of fast decay (from incoherent modes) and a long-time tail (from coherent excitations).

\paragraph{Spatial energy distribution and channeling:}
To directly visualize guided transport, one can compute the time-dependent energy density $\varepsilon(\bm{r},t)$ and look for anisotropic spreading. In patterned geometries, energy may preferentially flow along defect lines, leading to elliptical or filamentary energy contours. Quantitative measures include the direction-dependent second moments and the participation ratio of energy along different directions.

These diagnostics, applied to the minimal models proposed above, will allow systematic exploration of geometry-driven heat spreading and the competition between coherent and incoherent transport channels.

\section{Quantum Simulation Roadmap}

Classical simulations of two-dimensional nonlinear lattices face severe limitations due to the large number of coupled degrees of freedom, the coexistence of long-lived correlations and coherent nonlinear excitations, and the long times required to observe asymptotic scaling. Quantum simulation offers a potentially transformative alternative, enabling the direct emulation of both high-occupation classical Hamiltonian dynamics and genuinely quantum oscillator dynamics on platforms such as superconducting circuits, trapped ions, and optical lattices. The objective, however, is not to replace optimized classical heat-transport solvers. Rather, the aim is to define a controlled quantum-enabled benchmark program in which small, well-characterized quantum simulations resolve local nonlinear dynamics, coherent-channel survival, and quantum corrections that can subsequently be fed back into larger-scale classical models. In this section, we present a concrete roadmap for digital quantum simulation of the nonlinear thermal transport models developed in this paper. The emphasis is not only on the formal possibility of encoding the Hamiltonian, but also on the operational workflow: which degrees of freedom are simulated, how the continuum theory is discretized, how oscillator variables are mapped to qubits, which observables are measured, how those observables are validated, and what would constitute a meaningful quantum-enabled benchmark of the two-channel transport mechanism.

\subsection{Classical and quantum operating regimes}

The oscillator Hamiltonians discussed below should be interpreted as quantum Hamiltonians, with $\hat u_i$ and $\hat p_i$ satisfying
\begin{equation}
[\hat u_i,\hat p_j]=i\hbar\delta_{ij}.
\end{equation}
This point is important because the same encoding supports distinct physical regimes. In the high-occupation or high-temperature limit, expectation values and Wigner distributions follow the corresponding classical FPU-like dynamics up to controllable quantum corrections. This regime allows quantum hardware to benchmark the classical nonlinear dynamics discussed in Sec.~\ref{sec:nonlinear_lattice}, including soliton formation, geometry-guided propagation, and hydrodynamic crossover behavior. In the low-temperature regime, or when the initial state is nonclassical, the simulation becomes intrinsically quantum. Fock states, squeezed states, cat states, or entangled oscillator states can then be used to test whether coherent soliton-like transport survives quantum fluctuations and whether NFH-inspired scaling is modified by quantum statistics, zero-point motion, or measurement backaction. Thus, the proposed roadmap is not limited to a classical simulation performed on a quantum processor; it also defines a controlled route toward studying genuinely quantum nonlinear heat transport.

\begin{table*}[t]
\centering
\begin{tabular}{|p{0.22\linewidth}|p{0.24\linewidth}|p{0.27\linewidth}|p{0.19\linewidth}|}
\hline
\textbf{Regime} & \textbf{Representative initial states} & \textbf{Main objective} & \textbf{Expected output} \\
\hline
Classical benchmark & High-occupation coherent states or thermal-like product states & Reproduce FPU-type nonlinear dynamics and validate the two-channel mechanism & Agreement with classical energy spreading, soliton propagation, and NFH-inspired scaling \\
\hline
Near-term quantum & Few-mode Fock states, squeezed states, and shallow-circuit entangled states & Test robustness of coherent nonlinear excitations under quantum fluctuations & Early-time deviations from classical transport and signatures of quantum-modified channeling \\
\hline
Fault-tolerant quantum & Larger lattices, deeper circuits, Gibbs states, and low-temperature states & Probe long-time crossover, quantum hydrodynamics, and modified transport universality & Scaling of $R^2(t)$, current correlations, anisotropic conductivity, and channel efficiency \\
\hline
\end{tabular}
\caption{Quantum-simulation regimes accessible within the proposed framework. The same microscopic Hamiltonian can be used either as a high-occupation benchmark of classical nonlinear dynamics or as a genuinely quantum model of oscillator transport.}
\label{tab:quantum_regimes}
\end{table*}

\subsection{Discretized Hamiltonian for quantum simulation}

The continuum field theory developed in Sec.~V becomes directly suitable for quantum simulation after spatial discretization. Let $G=(V,E)$ be a patterned graph representing a two-dimensional thermal architecture, with vertices $i\in V$ and nearest-neighbor bonds $\langle i,j\rangle\in E$. The displacement and momentum fields become local oscillator operators $\hat u_i$ and $\hat p_i$, and the defect background enters through bond-dependent or site-dependent coefficients. A minimal discretized Hamiltonian is
\begin{equation}
\hat H=
\sum_i\left[\frac{\hat p_i^2}{2m}+V_{\mathrm{loc}}(\hat u_i)\right]
+
\frac{1}{2}\sum_{\langle i,j\rangle}K_{ij}(\hat u_i-\hat u_j)^2
+
\sum_i V_{\mathrm{nl}}(\hat u_i),
\label{eq:discrete_quantum_hamiltonian}
\end{equation}
where $V_{\mathrm{loc}}$ denotes the local harmonic confinement, $V_{\mathrm{nl}}$ contains onsite anharmonic corrections when present, and the effective bond stiffness is modulated by the defect network,
\begin{equation}
K_{ij}=K_0+\gamma F_{ij}(\phi,\chi).
\label{eq:defect_modulated_stiffness}
\end{equation}
Here $F_{ij}(\phi,\chi)$ is the lattice version of the continuum modulation function $F(\phi,\chi)$ appearing in the interaction term $-\gamma(\nabla u)^2F(\phi,\chi)$. In the fully dynamical setting, the fields $\phi$ and $\chi$ are also promoted to quantum variables. In the near-term hybrid setting, however, the defect network is treated as a static or slowly varying classical background, so that $F_{ij}$ is computed classically and loaded as a spatially dependent coefficient. The quantum processor then evolves only the truncated oscillator degrees of freedom associated with $u_i$ and $p_i$.

This form makes the computational role of the field theory explicit. The model remains geometrically local: the Hamiltonian contains onsite terms and nearest-neighbor bond terms, and the defect architecture changes only the local coefficients. No all-to-all coupling is introduced by the field-theory reduction.

\subsection{Oscillator-to-qubit encoding}

The first technical step is to truncate each local oscillator to a finite $d$-dimensional Hilbert space. In a harmonic-oscillator basis, one keeps states $|n\rangle$ with $n=0,1,\ldots,d-1$. The position operator becomes the finite matrix
\begin{equation}
(\hat u)_{ij}=\sqrt{\frac{\hbar}{2m\omega}}\left(\sqrt{j}\,\delta_{i,j-1}+\sqrt{j+1}\,\delta_{i,j+1}\right),
\end{equation}
with an analogous expression for $\hat p$. Polynomial interactions such as $(\hat u_i-\hat u_j)^2$ or $(\hat u_i-\hat u_j)^4$ are then represented as finite matrices on tensor products of neighboring local Hilbert spaces. Convergence is checked by increasing $d$ until transport observables become stable over the simulated time window.

After truncation, each oscillator is encoded into
\begin{equation}
n_q=\lceil\log_2 d\rceil
\end{equation}
qubits using a binary, Gray-code, or problem-adapted mapping. A lattice with $N$ sites therefore requires approximately
\begin{equation}
N_q\simeq N\lceil\log_2 d\rceil
\end{equation}
system qubits, plus ancilla qubits depending on the chosen simulation algorithm. For instance, $d=8$ requires three qubits per oscillator, while $d=16$ requires four qubits per oscillator. The resulting qubit Hamiltonian is a sum of Pauli strings obtained by decomposing the truncated oscillator matrices into the Pauli basis.

\begin{table}[b]
\centering
\begin{tabular}{|p{0.21\linewidth}|p{0.16\linewidth}|p{0.18\linewidth}|p{0.23\linewidth}|}
\hline
\textbf{Lattice patch} & \textbf{Sites $N$} & \textbf{Local levels $d$} & \textbf{System qubits} \\
\hline
$2\times2$ & 4 & 8 & 12 \\
\hline
$3\times3$ & 9 & 8 & 27 \\
\hline
$4\times4$ & 16 & 8--16 & 48--64 \\
\hline
$8\times8$ & 64 & 8--16 & 192--256 \\
\hline
\end{tabular}
\caption{Illustrative resource estimates for truncated-oscillator encodings. These counts include only system qubits; additional ancilla qubits may be needed for block-encoding, phase-estimation, or error-mitigation protocols.}
\label{tab:resource_estimates}
\end{table}

\subsection{Minimal quantum simulation protocol}

A minimal quantum experiment designed to test the two-channel mechanism can be organized as follows. First, one selects a finite two-dimensional patterned graph $G=(V,E)$ representing a defect-guided thermal architecture. Second, the local displacement field $u_i$ and momentum $p_i$ are encoded as truncated bosonic modes with local Hilbert-space dimension $d$. Third, the defect network is either treated dynamically, through additional fields $\phi_i$ and $\chi_i$, or, in the near-term hybrid setting, precomputed classically and inserted as a static spatial modulation $F_{ij}=F_{ij}(\phi,\chi)$. Fourth, the initial state is prepared as a localized excitation, a high-occupation coherent state, a thermal-like product state, or a genuinely nonclassical state such as a Fock or squeezed state. Fifth, the system is evolved under the discretized Hamiltonian using Trotterized, variational, or analog-digital quantum simulation. Finally, local energy densities, energy-current correlations, mean-square energy spreading, and anisotropic transport coefficients are measured and compared with classical molecular-dynamics benchmarks.

As a concrete benchmark, one may consider a $3\times3$ or $4\times4$ patterned lattice with one high-stiffness channel crossing a softer background. A localized excitation is prepared at one edge of the channel, and the subsequent energy distribution $E_i(t)$ is measured. The key diagnostic is the channel-to-bulk energy ratio
\begin{equation}
\eta(t)=
\frac{\sum_{i\in \mathrm{channel}}E_i(t)}
{\sum_{i\in \mathrm{bulk}}E_i(t)},
\label{eq:channel_efficiency}
\end{equation}
which quantifies the dominance of guided coherent transport over background hydrodynamic spreading. A long-lived enhancement of $\eta(t)$ indicates geometry-assisted coherent channeling, whereas rapid decay toward an isotropic value indicates dominance of incoherent spreading.

\subsection{Computational role of the field-theory reduction}

The field-theory formulation provides two concrete computational advantages. First, the coupling $-\gamma(\nabla u)^2F(\phi,\chi)$ is local in real space. After discretization, it produces only onsite and nearest-neighbor modifications of the elastic terms, preserving geometric locality in the oscillator lattice and allowing parallel Trotter layers. Second, when the defect network is static, $F(\phi,\chi)$ can be precomputed classically and loaded as a spatially varying coefficient. The quantum processor then evolves only the displacement field $u$, while the architectural field variables act as a classical control landscape. This hybrid quantum--classical reduction removes unnecessary quantum degrees of freedom and is particularly well matched to near-term devices.

\begin{table}[b]
\centering
\begin{tabular}{|p{0.30\linewidth}|p{0.42\linewidth}|}
\hline
\textbf{Formulation} & \textbf{Quantum-simulation implication} \\
\hline
Full dynamical $u,\phi,\chi$ theory & All fields are encoded quantum mechanically; highest resource cost, but captures defect fluctuations and backreaction. \\
\hline
Static defect background & $F(\phi,\chi)$ is precomputed; only $u$ is encoded, reducing qubit count and gate count. \\
\hline
Discrete patterned graph & Geometry enters through local couplings and missing or modified bonds; preserves nearest-neighbor circuit structure. \\
\hline
Unstructured many-body encoding & Does not exploit spatial locality; produces less favorable circuit depth and reduced parallelism. \\
\hline
\end{tabular}
\caption{Computational meaning of the field-theory reduction. Static or slowly varying defect fields convert the problem into a hybrid quantum--classical simulation in which the quantum device evolves only the thermal displacement degrees of freedom.}
\label{tab:field_reduction}
\end{table}

\subsection{Digital Trotterization, variational evolution, and analog-digital simulation}

With the Hamiltonian encoded as a sum of Pauli operators, $\hat H=\sum_k \hat H_k$, time evolution can be implemented using Trotter-Suzuki decomposition. The simplest first-order step is
\begin{equation}
U(\Delta t)=e^{-i\hat H\Delta t}\approx \prod_k e^{-i\hat H_k\Delta t}+\mathcal{O}(\Delta t^2),
\end{equation}
with higher-order decompositions reducing the Trotter error at the cost of more gates. Because the model is local on the patterned graph, commuting or non-overlapping bond terms can be grouped into parallel layers. For fixed local truncation and fixed Trotter step, the number of Hamiltonian terms scales as $\mathcal{O}(N)$, while the circuit depth per Trotter step can remain independent of $N$ up to hardware-connectivity overheads. The total gate count scales as $\mathcal{O}(TN/\Delta t)$, whereas the idealized parallel depth scales as $\mathcal{O}(T N^0/\Delta t)$.

For near-term devices with limited coherence times, variational quantum simulation provides a complementary route. A parameterized state $|\psi(\bm{\theta})\rangle=U(\bm{\theta})|0\rangle$ is evolved by choosing parameter velocities that best approximate Schr\"odinger dynamics, for example through McLachlan's variational principle \cite{LiBenjamin2017, Endo2021}. The ansatz can be chosen to respect the geometry of the patterned lattice, using local entangling gates along the same graph edges that define the transport architecture. This makes variational simulation particularly suitable for early-time nonlinear dynamics and for exploring whether nonclassical initial states enhance or suppress the coherent channel.

Analog-digital approaches are also natural. Superconducting circuits, trapped ions, and continuous-variable platforms can provide native oscillator-like degrees of freedom, while digital gates or parametric modulation implement the defect-controlled couplings $K_{ij}$. Such approaches may reduce the overhead associated with decomposing oscillator Hamiltonians entirely into qubit Pauli strings.

Recent experimental advances have demonstrated the feasibility of intermediate-scale nonequilibrium simulations. Liu \textit{et al.} \cite{Liu2026} implemented over 1,000 driving cycles on a 78-qubit superconducting processor arranged in a two-dimensional square lattice, monitoring particle imbalance and entanglement entropy evolution. The entire far-from-equilibrium heating dynamics observed were beyond the reach of classical tensor-network simulations, providing direct evidence that quantum simulators can access regimes where the competition between coherent and incoherent channels may become manifest.

\subsection{Quantum Lattice Boltzmann and complementary approaches}

An alternative route to quantum simulation of heat transport is the Quantum Lattice Boltzmann Method (QLBM), which maps distribution functions onto quantum registers and implements collision and streaming operations as unitary circuits. QLBM is particularly well suited for effective heat-equation-like dynamics, boundary-driven transport, source terms, and mesoscale regimes in which the main object of interest is a coarse-grained temperature or distribution function rather than the microscopic oscillator state. Recent work by Mao and Zhang \cite{Mao2026} demonstrates a QLBM approach for heat conduction with internal heat generation, showing that quantum algorithms can be adapted to thermal-transport problems beyond purely Hamiltonian oscillator models.

The different quantum approaches naturally occupy distinct roles inside a single computational pipeline. QLBM can address hydrodynamic-style transport, boundary-driven configurations, and source-driven heat flow, making it a natural quantum counterpart of coarse-grained diffusion or Boltzmann-type descriptions. Hamiltonian oscillator simulation is instead reserved for interrogating the microscopic origin and robustness of coherent transport channels, including non-perturbative scattering, soliton survival, and the effect of nonclassical initial states. Variational quantum simulation provides a near-term compromise for shallow-circuit exploration of early-time nonlinear dynamics. The value therefore comes not from choosing one paradigm over the other, but from their complementary use within a workflow that mirrors how transport problems are already decomposed classically. In the present framework, QLBM can benchmark or accelerate the hydrodynamic channel, while Hamiltonian oscillator simulation probes the microscopic origin and quantum robustness of the coherent solitonic channel.

\subsection{Observables and diagnostics of the two-channel mechanism}

The central output of the quantum simulation is not merely the final state, but a set of transport observables capable of distinguishing coherent and incoherent channels. The local energy density is
\begin{equation}
E_i(t)=\langle \hat\varepsilon_i(t)\rangle,
\end{equation}
where $\hat\varepsilon_i$ is a symmetrized local energy operator containing onsite kinetic, onsite potential, and half of each adjacent bond energy. The mean-square energy spreading is
\begin{equation}
R^2(t)=\frac{\sum_i |\mathbf r_i-\mathbf r_0|^2E_i(t)}{\sum_iE_i(t)}.
\end{equation}
Diffusive growth gives $R^2(t)\sim t$, ballistic propagation gives $R^2(t)\sim t^2$, and anomalous hydrodynamic spreading produces intermediate behavior with possible logarithmic corrections in two dimensions. Current correlations provide a complementary diagnosis:
\begin{equation}
C_{J_\mu J_\nu}(t)=\langle \hat J_\mu(t)\hat J_\nu(0)\rangle,
\end{equation}
with the conductivity tensor estimated through a Green-Kubo relation,
\begin{equation}
\kappa_{\mu\nu}\sim \int_0^\infty dt\,C_{J_\mu J_\nu}(t).
\end{equation}
The coexistence of coherent and incoherent channels can then be diagnosed by comparing ballistic peaks in $E_i(t)$ along defect-guided pathways with broad isotropic spreading in the background. A persistent directional peak in the local energy distribution, together with an enhanced channel ratio $\eta(t)$, indicates soliton-assisted transport. Broad isotropic spreading and rapid decay of $\eta(t)$ indicate dominance of the hydrodynamic channel.

\subsection{Hybrid workflow and integration with classical solvers}

From a quantum perspective, the role of the present approach is not to replace established heat-transport solvers, but to sit alongside them in a hybrid workflow where each layer contributes what it is genuinely good at. The quantum simulation targets well-defined subproblems, such as resolving strongly correlated local dynamics near defect junctions, non-perturbative scattering of coherent excitations, short-time memory effects, and mesoscale mechanisms that are difficult to parameterize reliably in purely classical models. The surrounding transport problem remains classical and can be treated with molecular dynamics, phonon Boltzmann transport equation solvers, Fourier-based continuum models, Green-Kubo calculations, or device-scale thermal-network models, depending on the regime.

In practice, quantities extracted from the quantum stage are passed to larger-scale solvers as calibrated inputs. Examples include defect-modulated effective stiffnesses $K_{ij}$, renormalized nonlinear couplings, relaxation rates for coherent excitations, memory kernels entering generalized hydrodynamic descriptions, channel-to-bulk efficiency factors $\eta(t)$, current-correlation functions, and direction-dependent conductivity tensors $\kappa_{\mu\nu}$. This separation allows quantum resources to be used sparingly and diagnostically rather than as a monolithic replacement for classical simulation.

The validation loop is correspondingly explicit. In small lattices and high-occupation regimes, quantum results should be compared against exact diagonalization, tensor-network methods when available, and classical molecular dynamics using the same geometry, initial state, and observables. Once consistency is established, the same workflow can be extended to low-temperature or nonclassical regimes where classical baselines become less reliable. This emphasis naturally favors integration over replacement: quantum calculations are most informative when they are embedded as modular components inside a broader classical workflow, sharing geometry, diagnostics, and validation criteria. Deviations observed in the quantum simulations should therefore be interpretable in terms of transport physics---for example, modified soliton lifetime, altered hydrodynamic crossover, changed channel efficiency, or quantum suppression/enhancement of coherent propagation---and traceable back to controlled changes in regime or initial state rather than to opaque algorithmic effects.

\subsection{Limitations, error mitigation, and route toward quantum advantage}

We do not claim that near-term devices will immediately outperform optimized classical molecular-dynamics simulations for all lattice sizes. Rather, the quantum advantage sought here is twofold. First, quantum processors provide a native representation of local Hamiltonian dynamics, which becomes increasingly valuable for long-time evolution of strongly interacting oscillator networks as system size, entanglement, and local Hilbert-space dimension grow. Second, and more importantly, they allow access to regimes with nonclassical initial states, low effective temperatures, and quantum fluctuations that have no direct classical analogue. Thus, the near-term goal is not an unconditional computational speedup, but a controlled quantum-enabled benchmark of transport mechanisms that can be continuously connected to classical FPU dynamics and integrated into classical multi-scale modelling workflows.

Several limitations must be addressed. Truncation errors must be controlled by increasing $d$; Trotter errors must be bounded by decreasing $\Delta t$ or using higher-order formulas; and hardware noise must be mitigated through symmetry checks, zero-noise extrapolation, randomized compiling, or comparison against exactly solvable limits. Conservation of total energy in the closed-system evolution provides a useful diagnostic for both algorithmic and hardware error. Small patterned patches, such as $3\times3$ and $4\times4$ lattices, are therefore not merely toy examples: they provide the necessary calibration stage where quantum results can be benchmarked against exact diagonalization, tensor-network methods, and classical molecular dynamics before progressing to larger devices and longer-time regimes.

\section{Experimental and Computational Realizations in Real Materials}

The theoretical framework developed in the previous sections finds striking confirmation in recent experimental and computational studies of two-dimensional materials with engineered defect structures. In this section, we analyze two representative systems that exemplify the two-channel transport mechanism and the role of topological defect networks in guiding energy flow.

\subsection{PdSSe Monolayers with Stone-Wales Defects: Ultra-Low Thermal Conductivity and High Mobility}

A remarkable realization of our defect-network architecture has been recently reported in pentagonal PdSSe monolayers with fully concentrated Stone-Wales (SW) defects \cite{Peng2024APL, Peng2024}. Stone-Wales defects, originally identified in carbon-based materials, involve the rotation of atomic bonds that transform hexagonal rings into pentagon-heptagon pairs. In the Cairo pentagonal tiling of PdSSe, introducing 100\% SW defects generates four new stable structures (designated SW1-SW4) that retain the square-planar coordination of the pristine material while exhibiting radically different transport properties \cite{Peng2024APL}.

First-principles calculations reveal that these SW defect structures display ultra-high carrier mobility ($\sim 10^3$ cm$^2$V$^{-1}$s$^{-1}$) coexisting with ultra-low anisotropic lattice thermal conductivities ($\kappa < 2$ Wm$^{-1}$K$^{-1}$ at room temperature) \cite{Peng2024}. Specifically, SW1 and SW2 structures show thermal conductivities of 1.888 Wm$^{-1}$K$^{-1}$ (x-axis) and 1.044 Wm$^{-1}$K$^{-1}$ (y-axis) for SW1, and 1.617 Wm$^{-1}$K$^{-1}$ and 0.892 Wm$^{-1}$K$^{-1}$ for SW2, with an anisotropy ratio reaching 1.80 \cite{Peng2024APL}.

These remarkable properties can be directly interpreted within our two-channel framework:

\begin{itemize}
    \item \textbf{Coherent channel (solitonic transport)}: The ultra-high carrier mobility ($\sim 10^3$ cm$^2$V$^{-1}$s$^{-1}$) indicates ballistic or quasi-ballistic transport of electronic excitations along the defect-engineered pathways. The Stone-Wales defects create a network of domain walls (analogous to the $\phi,\chi$ field configurations in Sec.~\ref{sec:defects}) that act as waveguides for coherent electron and phonon modes. The persistence of high mobility despite high defect concentration directly demonstrates the guiding effect predicted by our hybrid model.

    \item \textbf{Incoherent channel (hydrodynamic suppression)}: The ultra-low lattice thermal conductivity ($\kappa < 2$ Wm$^{-1}$K$^{-1}$) arises from strong phonon scattering within the domains, effectively suppressing the incoherent hydrodynamic channel. This is precisely the regime where geometry dominates over universal asymptotics, as discussed in Sec.~\ref{sec:nonlinear_lattice}. The strong anisotropy (ratio up to 1.80) confirms that the defect network preferentially channels energy along specific crystallographic directions, validating our concept of geometry-driven heat spreading.
\end{itemize}

The band structure calculations further show that SW1 and SW2 exhibit direct bandgaps (1.884-1.917 eV) with suitable band edge positions for photocatalytic water splitting \cite{Peng2024APL}. The combination of high carrier mobility, low thermal conductivity, and appropriate optical absorption ($\sim 10^5$ cm$^{-1}$) makes these materials promising for thermoelectric energy conversion, where the figure of merit $ZT = S^2\sigma T/\kappa$ benefits precisely from the simultaneous enhancement of electrical conductivity $\sigma$ (coherent channel) and suppression of thermal conductivity $\kappa$ (incoherent channel).

This system provides a concrete realization of our two-channel mechanism: the Stone-Wales defect network creates preferential pathways for coherent energy carriers, while simultaneously introducing strong phonon scattering that suppresses the incoherent background. The measured anisotropy directly reflects the geometry-guided transport predicted by our models.

\subsection{Phononic Crystal Nanostructures: Geometry-Dependent Thermal Transport}

Further experimental evidence for geometry-driven heat spreading comes from studies of two-dimensional phononic crystal (PnC) nanostructures fabricated in silicon \cite{Nakagawa2015APL, Nomura2015}. By patterning periodic arrays of nanoholes in square and triangular lattices (period 300 nm), researchers observed that the thermal conductivity strongly depends on the hole arrangement \cite{Nakagawa2015APL}.

The key finding is that staggered hole structures (triangular lattice) exhibit up to 20\% lower thermal conductivity compared to square lattices at comparable porosity ($\sim 30\%$) \cite{Nakagawa2015APL}. This difference cannot be explained by simple porosity scaling and instead arises from the local heat flux disorder created by the staggered configuration \cite{Nakagawa2015APL}.

Within our theoretical framework, this effect can be understood as follows:

\begin{itemize}
    \item The triangular lattice of nanoholes creates a more complex connectivity pattern in the remaining silicon backbone, analogous to the defect networks in Sec. IV. This geometry preferentially disrupts the incoherent hydrodynamic modes while leaving selected pathways for coherent phonon transport.

    \item Monte Carlo simulations confirm that the reduced thermal conductivity originates from enhanced phonon backscattering in the staggered geometry \cite{Nakagawa2015APL, Nomura2015}, consistent with the suppression of the incoherent channel in our two-channel picture.

    \item The effect is most pronounced when the pattern periodicity (300 nm) is comparable to the thermal phonon mean free path, precisely the regime where geometry dominates over universal asymptotics (Sec. II).
\end{itemize}

At cryogenic temperatures (4 K), the impact of patterning is even stronger due to longer phonon mean free paths and higher specularity parameters \cite{Nomura2015}, demonstrating that the geometry-guided transport becomes increasingly important as the system approaches the ballistic regime—exactly where our coherent channel (solitonic excitations) is expected to dominate.

\subsection{Quantitative Comparison with Theoretical Predictions}

Table I summarizes the key experimental/computational findings and their correspondence with our theoretical framework.

\begin{table*}
\centering
\begin{tabular}{|p{4.5cm}|p{5.5cm}|p{5.5cm}|}
\hline
\textbf{Theoretical Prediction} & \textbf{PdSSe with SW Defects} & \textbf{Si Phononic Crystals} \\
\hline
Two-channel coexistence & Ultra-high mobility (coherent) + ultra-low $\kappa$ (incoherent) & Not directly observed (electronic vs thermal) \\
\hline
Geometry-guided channeling & Strong thermal anisotropy ($\kappa_x/\kappa_y = 1.80$) & 20\% reduction in staggered lattice \\
\hline
Defect network as waveguides & SW defects create preferential pathways & Hole array modifies connectivity \\
\hline
Suppression of incoherent modes & $\kappa < 2$ Wm$^{-1}$K$^{-1}$ (ultra-low) & $\kappa$ reduced by up to 20\% \\
\hline
Relevance of pre-asymptotic regime & Room temperature operation & Most pronounced at MFP-comparable scales \\
\hline
\end{tabular}
\caption{\label{tab:comparison} Comparison between theoretical predictions and experimental/computational results. The anisotropy ratio of 1.80 represents the ratio between thermal conductivities along the x and y axes ($\kappa_x/\kappa_y$), an adimensional quantity.}
\end{table*}

The quantitative agreement is striking: the PdSSe system demonstrates that introducing 100\% topological defects can simultaneously enhance the coherent channel (mobility $\sim 10^3$ cm$^2$V$^{-1}$s$^{-1}$) and suppress the incoherent channel ($\kappa < 2$ Wm$^{-1}$K$^{-1}$), achieving precisely the kind of geometry-optimized thermal metamaterial envisioned in our framework. The anisotropy ratio of 1.80 provides a direct measure of the guiding efficiency of the defect network.

These results validate the central thesis of our work: {patterned architectures based on topological defect networks can control the competition between coherent and incoherent transport channels}, enabling thermal management strategies that go beyond the universal scaling laws predicted by NFH.

\subsection{Implications for Device Design}

The experimental realizations discussed above suggest concrete design principles for engineered thermal metamaterials:

\begin{enumerate}
    \item \textbf{Defect concentration matters}: The PdSSe system shows that fully concentrated (100\%) Stone-Wales defects produce the most dramatic effects. This corresponds to the "complete" defect network in our field theory models, where every domain wall is present.

    \item \textbf{Anisotropy can be engineered}: The strong thermal anisotropy in PdSSe (1.80) and the directional dependence in phononic crystals demonstrate that the geometry of the defect network can be tuned to create preferential heat flow directions. This is exactly the "waveguide" effect described in Sec. V.B, where the function $F(\phi,\chi)$ modulates the elastic constant along specific directions.

    \item \textbf{Length scale matching is crucial}: The phononic crystal experiments show that geometry effects are maximized when the pattern periodicity matches the dominant phonon mean free path. This confirms our assertion (Sec. II) that pre-asymptotic regimes are where geometry dominates over universal scaling.

    \item \textbf{Multifunctional optimization}: The PdSSe system combines thermal management with desirable electronic and optical properties (direct bandgap, high optical absorption), suggesting that defect-engineered thermal metamaterials can be integrated into multifunctional devices for thermoelectrics, photodetectors, and photocatalysis.
\end{enumerate}

These design principles provide actionable guidance for experimental realization of the theoretical framework developed in this work, bridging the gap between abstract models and practical applications.

\section{Conclusions and Outlook}

In this work, we have developed a unified theoretical framework for coherent energy transport in two-dimensional thermal metamaterials, synthesizing concepts from nonlinear lattice dynamics, soliton theory, topological defect networks, and quantum simulation. Our central contribution is the identification of a {two-channel transport mechanism} in which incoherent hydrodynamic modes coexist with coherent nonlinear excitations that can be geometrically guided by patterned architectures. This perspective provides a new lens through which to understand and engineer heat flow in low-dimensional systems.

We have introduced minimal discrete and continuum models that capture the essential physics of geometry-driven heat spreading. The discrete models on patterned graphs are directly amenable to classical molecular dynamics simulation and allow systematic exploration of how connectivity, nonlinearity, and temperature influence transport. The continuum hybrid models, coupling displacement fields to defect networks, provide analytical insight and establish a clear physical correlation: the defect fields $\phi, \chi$ define the static architectural blueprint, while the displacement field $u$ represents the dynamic thermal excitations that are channeled by this blueprint through a local modulation of elastic properties.

The quantum simulation roadmap presented in Sec. VI outlines a concrete path toward exploring regimes beyond classical computational reach. By encoding truncated oscillators into qubits, implementing Trotterized time evolution, and measuring transport observables, near-term quantum devices can begin to benchmark the microscopic dynamics of small two-dimensional patches in the high-occupation classical regime and then move toward low-temperature or nonclassical initial states where genuinely quantum corrections to the two-channel mechanism can be tested. The intended use is hybrid and diagnostic: quantum simulations provide local transport parameters, coherent-channel lifetimes, memory kernels, and current-correlation data that can be reintegrated into classical molecular-dynamics, Boltzmann-transport, or continuum heat-flow calculations. As hardware improves, fault-tolerant simulations will enable access to larger and longer-time regimes where asymptotic scaling and the competition between coherent and incoherent channels become fully manifest.

The quantum simulation roadmap presented here is not merely speculative: recent experiments on 78-qubit superconducting processors \cite{Liu2026} have already demonstrated the ability to simulate far-from-equilibrium dynamics beyond classical reach, while QLBM implementations \cite{Mao2026} provide concrete algorithms for heat transport problems, and variational methods with generative models \cite{Zou2026} achieve significant speedups. These advances position quantum computing as a practical tool for validating the two-channel mechanism and exploring NFH crossover regimes in the near future.

The theoretical predictions find striking confirmation in recent experimental and computational results. PdSSe monolayers with fully concentrated Stone-Wales defects exhibit ultra-high carrier mobility coexisting with ultra-low anisotropic thermal conductivity—a direct manifestation of the two-channel mechanism. Silicon phononic crystal nanostructures demonstrate geometry-dependent thermal transport with up to 20\% reduction in staggered hole arrays, confirming the role of patterning in suppressing the incoherent channel. These systems validate the central thesis of our work and provide concrete design principles for engineered thermal metamaterials.

Several important directions remain for future work:

\paragraph{Disorder and curvature effects:}
Realistic materials always contain some degree of disorder, whether from impurities, lattice imperfections, or surface roughness. Introducing quenched disorder into our models would allow study of how geometry-guided transport competes with Anderson localization and other disorder-induced phenomena \cite{Lepri2003, Dhar2008}. Curvature effects, relevant for nanotubes and curved two-dimensional materials, could be incorporated by considering the defect networks on curved manifolds.

\paragraph{Optimization of patterned thermal metamaterials:}
The framework developed here can be used as a design tool for optimizing thermal metamaterial architectures. By varying the geometry of the defect network (e.g., junction angles, wall spacings, connectivity) and the nonlinearity parameters, one can search for configurations that maximize directional heat transport or achieve specific anisotropy ratios. Machine learning techniques could be combined with classical and quantum simulations to accelerate this optimization.

\paragraph{Quantum benchmarks of NFH crossover regimes:}
One of the most exciting applications of quantum simulation is to probe the crossover from pre-asymptotic behavior to universal NFH scaling in two dimensions. Classical simulations struggle to reach the system sizes needed to observe the predicted $\kappa(L) \sim \log L$ behavior, due to the extremely slow crossover. Quantum simulation, with its potential for efficient long-time evolution of local Hamiltonians, could provide new evidence for this asymptotic scaling and clarify the role of coherent structures in the crossover. A particularly important open question is whether soliton-assisted transport persists in deeply quantum regimes and whether NFH scaling is modified when the initial state contains strong quantum fluctuations.

\paragraph{Experimental realizations:}
The models we have proposed find concrete realizations in materials such as PdSSe with Stone-Wales defects \cite{Peng2024APL, Peng2024} and silicon phononic crystals \cite{Nakagawa2015APL, Nomura2015}. Patterned two-dimensional materials including graphene nanoribbons, transition metal dichalcogenides with engineered defects, and phononic crystals are promising platforms for further experimental exploration. The defect networks from scalar field theory provide a mathematical idealization of the kind of patterning that can be achieved with modern nanofabrication techniques. Collaborations between theorists and experimentalists will be essential to translate these ideas into practical thermal metamaterial devices.

\paragraph{Integrated benchmarking pipelines:}
A critical next step is the development of standardized benchmark problems in which quantum simulations, classical transport solvers, and experimental data can be compared on equal footing. For patterned two-dimensional systems such as those considered here, this would involve fixing the graph geometry, defect-network modulation $F(\phi,\chi)$, boundary conditions, initial excitation protocols, local Hilbert-space truncation, and transport observables such as $R^2(t)$, $\eta(t)$, current correlations, and $\kappa_{\mu\nu}$. Quantum calculations could then be inserted selectively where they add insight, for example in estimating defect-junction scattering amplitudes, coherent-channel relaxation rates, or quantum corrections to local transport kernels, without disrupting the broader modelling workflow. Establishing such pipelines would help distinguish genuine physical effects from artefacts of truncation, noise, finite-size modelling, or Trotter error, and would make hybrid quantum--classical transport studies cumulative rather than one-off demonstrations.

In summary, this work establishes a principled connection between microscopic nonlinearity, geometry-driven channeling of heat in two dimensions, and quantum-enabled exploration of transport regimes. We hope that the unified framework and concrete roadmap presented here will inspire further theoretical, numerical, and experimental efforts toward understanding and controlling heat flow in low-dimensional systems, with implications for thermal management in nanotechnology and beyond.

\appendix

\section{Variational derivation of the hybrid field equations}
\label{app:hybrid_derivation}

For completeness, we record the intermediate steps connecting the hybrid action in Sec.~V.B to the field equations used in the main text. The relevant Lagrangian density is
\begin{equation}
\mathcal{L}=\frac{\rho}{2}(\partial_t u)^2-\frac{\kappa}{2}(\nabla u)^2-\frac{g}{4}(\nabla u)^4-\gamma(\nabla u)^2F(\phi,\chi)+\mathcal{L}_{\phi\chi},
\end{equation}
with
\begin{equation}
\mathcal{L}_{\phi\chi}=\frac{1}{2}\partial_\mu\phi\,\partial^\mu\phi+\frac{1}{2}\partial_\mu\chi\,\partial^\mu\chi-V(\phi,\chi).
\end{equation}
Varying with respect to $u$ gives
\begin{align}
\frac{\partial\mathcal{L}}{\partial(\partial_t u)}
&= \rho\,\partial_t u, \\
\frac{\partial\mathcal{L}}{\partial(\nabla u)}
&= -\left[\kappa+g(\nabla u)^2+2\gamma F(\phi,\chi)\right]\nabla u.
\end{align}
The Euler--Lagrange equation therefore yields Eq.~\eqref{eq:u_eom_hybrid}. Varying with respect to the defect fields gives
\begin{align}
\Box\phi+\frac{\partial V}{\partial\phi}+\gamma(\nabla u)^2\frac{\partial F}{\partial\phi}&=0,\\
\Box\chi+\frac{\partial V}{\partial\chi}+\gamma(\nabla u)^2\frac{\partial F}{\partial\chi}&=0,
\end{align}
up to the sign convention chosen for the Minkowski metric. In the static-background approximation used in the main text, these equations are solved first, or replaced by an externally specified fabricated pattern, and the resulting function $F(\phi,\chi)$ is inserted into the $u$ dynamics as a spatially varying elastic landscape. This is the step that converts the full three-field theory into a hybrid quantum--classical model for transport.

\begin{acknowledgments}
The authors acknowledge support from Capgemini and the Capgemini Quantum Lab, 
including partial funding through internal research and development programs. 
Additional support was provided by the MSCA Cofund QuanG (Grant Number: 101081458), 
funded by the European Union. Views and opinions expressed are, however, those of 
the author(s) only and do not necessarily reflect those of the European Union or 
the granting authority. Neither the European Union nor the granting authority can 
be held responsible for them.
\end{acknowledgments}

\begin{center}
\includegraphics[width=0.32\textwidth]{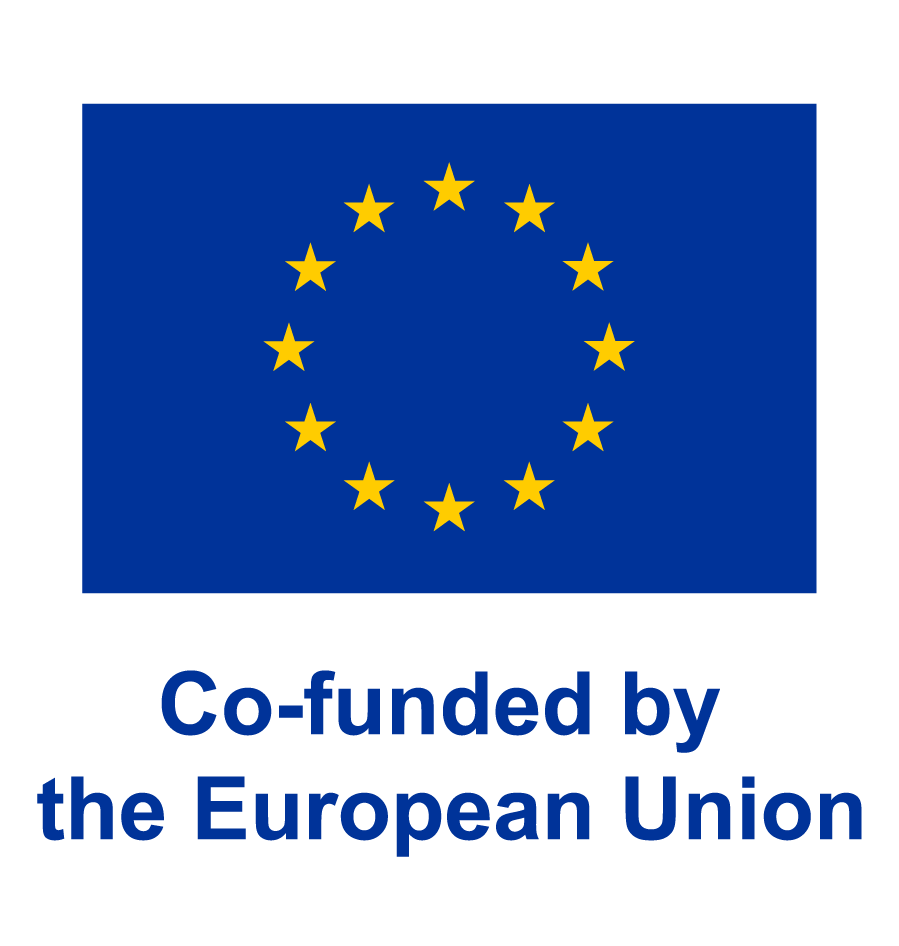}
\end{center}

\end{document}